\documentclass[lettersize,journal]{IEEEtran}
\usepackage{pgfplots}
\usepackage{pgfplotstable}
\pgfplotsset{compat=newest}
\usepackage{amsmath,amssymb,amsfonts}
\usepackage{algorithmic}
\usepackage{graphicx}
\usepackage{textcomp}
\usepackage{xcolor}
\usepackage{ulem}
\usepackage{multirow}
\usepackage[font=footnotesize]{caption}
\usepackage{subcaption}
\usepackage{url}  %
\usepackage{xurl}  %
\usepackage{cite}
\usepackage{amsmath}
\usepackage{bm}  %
\usepackage{tikz}
\usetikzlibrary{arrows.meta,positioning,calc,shapes.geometric,fit,backgrounds}
\definecolor{paperblue}{RGB}{43,92,138}
\definecolor{paperbluefill}{RGB}{232,241,250}
\definecolor{papergreen}{RGB}{48,120,96}
\definecolor{papergreenfill}{RGB}{232,246,241}
\definecolor{paperpurple}{RGB}{112,84,153}
\definecolor{paperpurplefill}{RGB}{242,237,248}
\definecolor{papergray}{RGB}{110,118,125}
\definecolor{papergrayfill}{RGB}{245,246,247}
\definecolor{paperline}{RGB}{65,70,75}
\newcommand{\solverfigwidth}{0.92\textwidth}
\begin{document}

\title{Evaluating a GPU-Enabled Solver for Scalable Unit Commitment in Large Power Grids}

\author{Hussein Sharadga, Mohamad Chehade, Javad Mohammadi,~\IEEEmembership{Senior Member,~IEEE,}%
\thanks{ \noindent \copyright~2026 IEEE. Personal use of this material is permitted. Permission from IEEE must be obtained for all other uses, in any current or future media, including reprinting/republishing this material for advertising or promotional purposes, creating new collective works, for resale or redistribution to servers or lists, or reuse of any copyrighted component of this work in other works. Manuscript received June 12, 2026.}

\thanks{Financial support for this research was provided by the US Advanced Research Projects Agency–Energy (ARPA-E) under grant number DE-AR0001646, and by the National Science Foundation under grant numbers 2442420 and 2313768.}
\thanks{Hussein Sharadga is with The University of Texas at Austin, Austin, TX 78712, USA, and also with the Department of Mechanical Engineering, Prairie View A\&M University, Prairie View, TX, USA (e-mail: hssharadga@tamu.edu).}
\thanks{Mohamad Chehade is with the Department of Civil, Architectural, and Environmental Engineering, The University of Texas at Austin, Austin, TX 78712, USA (e-mail: mf36244@my.utexas.edu).}
\thanks{Javad Mohammadi is with the Department of Civil, Architectural, and Environmental Engineering, The University of Texas at Austin, Austin, TX 78712, USA (e-mail: javadm@utexas.edu).}
}

\markboth{Accepted for publication in IEEE Transactions on Industry Applications}%
{Sharadga \MakeLowercase{\textit{et al.}}: Evaluating a GPU-Enabled Solver for Scalable Unit Commitment in Large Power Grids}

\maketitle

\begin{abstract}
This paper evaluates a GPU-based solver for large-scale unit commitment (UC) problems. The solver uses GPU acceleration to efficiently solve relaxed UC subproblems, enabling faster generation of objective bounds and improving the convergence of crossover procedure compared with CPU-only optimization approaches. The solver is evaluated on large-scale power grids with 4,224, 6,049, and 6,717 buses under extended scheduling horizons using a detailed UC formulation that includes extensive operational constraints, resulting in high computational complexity with millions of variables and constraints. Results from 116 operating scenarios demonstrate substantial speedups while maintaining high solution quality. A concurrent CPU–GPU framework is also implemented, in which CPU-based and GPU-based solvers run simultaneously and computation terminates upon convergence of either solver, thereby leveraging the complementary strengths of both architectures.

\end{abstract}

\begin{IEEEkeywords}
GPU-based optimization, unit commitment, large-scale power systems, power system optimization, parallel computing.
\end{IEEEkeywords}

\section{Introduction}
\subsection{Motivation}

\IEEEPARstart{T}{he} electric power grid is a large-scale infrastructure that requires efficient and reliable operational strategies \cite{TPEC26, b1}. As modern power systems grow in size, interconnection, and operational complexity, reliability and resilience requirements pose significant computational and modeling challenges. To address these challenges, the U.S. Department of Energy (DOE) has promoted the development of advanced optimization and high-performance computing techniques for next-generation grid operations, exemplified by the Grid Optimization (GO) Challenges \cite{b2}.

Optimal power flow (OPF) is a core optimization framework for power system operation, determining generation dispatch, load allocation, energy storage schedules, and associated control variables \cite{b3,SCOPF-Javad}. To enhance operational security, security-constrained OPF (SCOPF) extends the OPF framework by explicitly incorporating contingency scenarios into the optimization process \cite{b4}.

Early SCOPF formulations emerged in an era of limited computational resources and thus relied heavily on simplified network representations, most notably DC optimal power flow (DCOPF) models that neglect voltage magnitudes and reactive power interactions \cite{b5}. Although advances in optimization methods and hardware have expanded computational capabilities, many widely used approaches still rely on these approximations, limiting their ability to address the full nonlinear AC power flow problem \cite{b2}. To bridge this gap, the DOE introduced the GO Challenges, which emphasize scalable solutions to the full AC power flow problem while simultaneously incorporating unit commitment decisions and security constraints \cite{b2}.

The GO Challenges target large-scale power system problems involving unit commitment, nonlinear AC network constraints, and other operational constraints, posing substantial computational challenges. To address these difficulties, existing research has explored a variety of methodological advances aimed at improving robustness and computational efficiency. Chevalier \cite{ref4} introduced a parallel Adam-based optimization scheme that exploits concurrency to accelerate gradient calculations and variable updates. Complementary work in \cite{ref3} demonstrated that explicitly modeling AC power flow equations alongside reserve and unit commitment constraints leads to stable and reliable solutions for the GO benchmarks.

Several \textcolor{black}{GO }studies have also focused on relaxation techniques and solver performance analysis. Bienstock and Villagra \cite{refX} proposed a linear outer-approximation of the AC optimal power flow problem using cutting-plane strategies. Holzer et al. \cite{refY} analyzed solver submissions to GO Challenge 3, providing insight into the scalability and effectiveness of different solution approaches.

In our earlier contributions \cite{TPEC, IEEE-TIA, NAPS2025}, we developed a decomposition-based solution framework for the GO Challenge. The method divides the problem into two stages (see Fig. 1 in \cite{IEEE-TIA}): (1) a DC approximation that resolves binary commitment decisions and enforces feasibility, followed by (2) a nonconvex AC optimization that determines continuous operating variables. In \cite{NAPS2025}, we further tightened the DC model (first stage) to accelerate convergence.

\subsection{Literature Review}

DCOPF is widely used as a tractable linear approximation for transmission-level power system optimization. Foundational studies have clarified the assumptions underlying the DC model and the resulting tradeoffs in accuracy and tractability \cite{Stott2009}. Building on this approximation, a substantial body of work has sought to improve the scalability of DCOPF and related security-constrained formulations through reformulation, constraint reduction, and learned structure\textcolor{black}{\cite{Hinojosa2018,Hinojosa2020,Xavier2024,Ardakani2013,Molzahn2018,Roald2019,Zhang2020,Ng2018,Deka2019}.}

One line of work represents transmission constraints using sensitivity-based quantities such as power transfer distribution factors (PTDFs), line outage distribution factors (LODFs), and related linear shift factors. Compared with angle-based formulations, these approaches can eliminate explicit voltage-angle variables and yield more compact preventive and corrective security-constrained DCOPF models \cite{Hinojosa2018,Hinojosa2020}. Related work has also shown that appropriately designed shift-factor formulations can reveal decomposable structure that is useful for decentralized or distributed optimization \cite{Xavier2024}.

Another line of work reduces computational burden by identifying the subset of transmission constraints that is most relevant in practice. This includes umbrella-constraint identification in DC-based security-constrained OPF \cite{Ardakani2013}, detection of redundant flow limits on parallel lines \cite{Molzahn2018}, analysis of implied constraint-satisfaction relationships under varying load conditions \cite{Roald2019}, and study of which line-flow constraints are active in realistic DCOPF instances \cite{Zhang2020}. These results collectively motivate screening, redundancy removal, and selective constraint enforcement. Data-driven approaches have exploited recurring structure in the DCOPF solution map. Prior work has shown that DCOPF solutions often concentrate on a relatively small family of active-constraint patterns \cite{Ng2018}, and that active sets can be predicted directly from operating conditions using neural-network classifiers \cite{Deka2019}.

Traditional DCOPF and UC relaxations are commonly solved using CPU-oriented methods such as simplex, dual-simplex, and interior-point algorithms. Although effective for many applications, these methods contain sequential operations and sparse matrix factorizations that limit fine-grained parallelism on CPU architectures. As power system models grow to millions of variables and constraints, computational and memory requirements increase substantially, motivating the development of GPU-oriented optimization methods that can better exploit massive parallelism. Recent studies \cite{GPU1, GPU2, GPU3} have therefore investigated GPU acceleration for DCOPF, which is commonly used as a relaxed intermediate model to generate bounds in UC optimization.

In \cite{GPU1}, DCOPF is formulated as both a linear program (LP) and a quadratic program (QP) and solved using gradient-based methods accelerated by GPUs. The approach was evaluated on IEEE 2000-, 4601-, and 10k-bus systems, achieving a 25\% speedup on the 10k-bus case and a maximum speedup of $5.43\times$ on the proposed networks. Similarly, \cite{GPU2} implemented a predictor--corrector interior-point method for DCOPF, with most computations performed on the GPU. The tested systems ranged from 30 to 3120 buses, with the largest case achieving approximately a $4\times$ speedup. 

While these results in \cite{GPU1, GPU2} demonstrate the effectiveness of GPU acceleration, both studies considered a limited number of scenarios and relatively modest problem sizes; even in the largest cases, all solvers completed within 15 seconds.

More recently, \cite{GPU3} addressed a larger DCOPF instance using a GPU-accelerated Halpern--Peaceman--Rachford algorithm, reporting a best-case speedup of $6\times$. However, the largest test case involved approximately 200,000 variables. In contrast, the DOE challenge described in \cite{b2} involves optimization problems with up to 5.773 million variables, representing nearly a $29\times$ increase in problem size compared to the largest instances considered in existing GPU-accelerated DCOPF studies.

\textcolor{black}{A summary comparison of existing GPU-based DCOPF studies and the proposed approach is provided in Table~\ref{tab:gpu_literature_comparison}.} Although prior work demonstrates the potential of GPU acceleration for DCOPF, existing studies remain constrained by problem scale and scenario complexity. This paper directly targets optimization problems at the scale and structural complexity of the DOE challenge in \cite{b2}, thereby evaluating GPU-enhanced optimization methods under conditions that reflect realistic, large-scale UC applications.

\begin{table*}[!t]
\caption{\textcolor{black}{Comparison of GPU-based DCOPF studies and the current GPU-accelerated UC approach}}
\label{tab:gpu_literature_comparison}
\centering
\begin{tabular}{llllll}
\hline
Reference & GPU Algorithm & Problem Size & Problem Formulation \\
\hline
\cite{GPU1} 
& Gradient-based 
& IEEE 10k-bus system, variable count not reported
& Simplified DCOPF and limited scenarios \\
\cite{GPU2} 
& Predictor-corrector IPM 
& Up to 3120-bus system, variable count not reported
& Smaller-scale DCOPF and limited scenarios \\
\cite{GPU3} 
& Halpern--Peaceman--Rachford  
& Up to 9214-bus system, $\approx$200,000 variables
& Does not consider full UC formulation \\
&&&and operational constraints \\
\textbf{This work} 
& PDHG with concurrent 
& Up to 6717-bus system, up to 5.773 million variables
& Large-scale UC using the full ARPA-E GO\\
&  CPU/GPU execution&&challenge formulation\\
\hline
\end{tabular}
\end{table*}

\subsection{Contribution}
This work evaluates a solution framework for large-scale UC optimization using GPU acceleration for computationally intensive power system applications. The framework is evaluated on the high-fidelity DOE Challenge 3 formulation \cite{GO3}, where extensive operational constraints produce million-scale optimization instances that motivate scalable GPU-based solution methods.

Building on insights from our earlier investigations \cite{TPEC, IEEE-TIA, NAPS2025}, we adapted a GPU-oriented approach in which the linear relaxation of the UC problem is solved using a Primal–Dual Hybrid Gradient (PDHG) method. The proposed framework accelerates convergence within crossover procedure. %
Performance is evaluated on benchmark systems with 4224, 6049, and 6717 buses, demonstrating significant computational efficiency gains without sacrificing solution quality. In addition, this work develops a concurrent CPU--GPU framework in which CPU-based and GPU-based solvers are executed simultaneously, and computation terminates when either solver converges. This hybrid strategy leverages the complementary strengths of both architectures and further reduces average and worst-case computation times.
\section{Problem Setup}
\subsection{Problem Formulation}
The unit commitment model considered in this paper is based on the specifications released by the U.S. Department of Energy’s Advanced Research Projects Agency–Energy (ARPA-E) for the GO Competition Challenge 3 \cite{GO3, go_data}. The resulting optimization problem is characterized by an unusually high level of modeling fidelity, with the principal formulation documented across 53 pages \cite{GO3} and supported by an additional 22 pages of auxiliary material. Rather than focusing on a small set of constraints, the formulation integrates extensive operational, network, and temporal considerations. These include voltage magnitude restrictions and nodal limits on active and reactive power, reserve requirements imposed at the zonal level, discrete device commitment and switching logic, minimum on/off duration constraints, bounds on cumulative start-up events over multiple scheduling intervals, ramping limitations, and inter-temporal energy constraints across the planning horizon. A schematic representation summarizing the structure of this formulation can be found in Figure 1 of \cite{TPEC}.
The number of optimization variables can exceed 5 million continuous variables and more than 100,000 binary variables, with over 5 million constraints. 

Despite the extensive structure and scale of the complete formulation, the optimization problem can be represented symbolically in the following compact form:

\begin{equation}
\begin{aligned}
\max_{x} \quad & z(x) \\
\text{s.t.} \quad
& G(x)=0 \\
& H(x)\le 0
\end{aligned}
\end{equation}

where $x$ collectively represents the large set of continuous and binary decision variables associated with network operation, device scheduling, commitment logic, reserve allocation, and temporal coupling constraints. The objective function $z(x)$ denotes the total market surplus over the scheduling horizon, which includes appropriately signed terms for consumer values, producer costs, device startup and shutdown costs, fixed operating costs, branch-limit overload penalties, reserve procurement costs, bus real and reactive power mismatch penalties, and zonal reserve shortfall penalties. The compact notation above is used solely for presentation purposes; the functions $G(\cdot)$ and $H(\cdot)$ implicitly represent the complete large-scale formulation described in the 53-page primary specification and the additional 22 pages of supporting material \cite{GO3}.

\subsection{Primal–Dual Hybrid Gradient Method}
Linear programming has traditionally been solved using two dominant classes of algorithms: simplex-based techniques and interior-point methods. These solvers have reached a high level of maturity and provide robust performance across many applications. Nevertheless, their reliance on dense linear algebra operations and matrix factorizations makes efficient deployment on graphics processing units (GPUs) nontrivial.

The PDHG algorithm provides a scalable, first-order alternative tailored to large optimization problems. When specialized for linear programs, it is commonly referred to as the Primal–Dual Linear Programming (PDLP) method \cite{app2021}. In contrast to simplex and interior-point approaches, PDHG employs gradient-driven updates and inexpensive operations, primarily sparse matrix–vector products, which align well with GPU architectures. The method is derived from a primal–dual saddle-point formulation, where primal and dual variables are updated simultaneously through iterative steps. Originally proposed by Chambolle and Pock \cite{Chambolle2011}, the framework was later extended by Applegate et al. \cite{app2021} through the introduction of adaptive step-size selection and restart strategies, leading to improved convergence behavior and computational efficiency.

The primal LP and its corresponding dual formulation are given by:
\begin{alignat}{3}
&\text{Primal LP:} \quad && \min_{x} \quad && c^\top x \\
&                         && ~~\text{s.t.} && Ax \le b, \; x \ge 0 \nonumber \\
&\text{Dual LP:} \quad   && \max_{y} \quad && b^\top y \\
&                         && ~~\text{s.t.} && A^\top y \ge c, \; y \ge 0 \nonumber
\end{alignat}
An equivalent saddle-point formulation of the primal–dual pair is given by:
\begin{equation}
\min_{x} \max_{y} \; L(x,y) = c^\top x - y^\top A x + b^\top y
\end{equation}
The PDHG method proceeds by iteratively updating the primal and dual variables until convergence:\begin{align}
x_{k+1} &= \mathrm{proj}_{\mathbb{R}^n_+}\Big(x_k + \tau A^\top y_k - \tau c \Big) \\
y_{k+1} &= y_k - \sigma A \big( 2 x_{k+1} - x_k \big) + \sigma b
\end{align}
The parameters $\tau$ and $\sigma$ correspond to the primal and dual step sizes. Applegate et al. \cite{app2021} further refined PDHG algorithm with adaptive step-size control and restart strategies, boosting its performance on large LP instances.

In \cite{app2021}, PDHG was generally slower than the simplex, dual simplex, and barrier methods, though in some cases it was only slower by a factor of two. PDHG outperformed these traditional approaches in just 2\% of the test cases, particularly on large-scale problems. In this study, the relaxed subproblem of unit commitment in large-scale power grids is identified as one that can benefit from PDHG with GPU acceleration, potentially achieving faster convergence than the simplex, dual Simplex, and barrier methods, with speedups potentially exceeding a factor of three (i.e., speedup $\approx 3\times$).

\begin{figure*}
    \centering
    \includegraphics[width=0.9\linewidth]{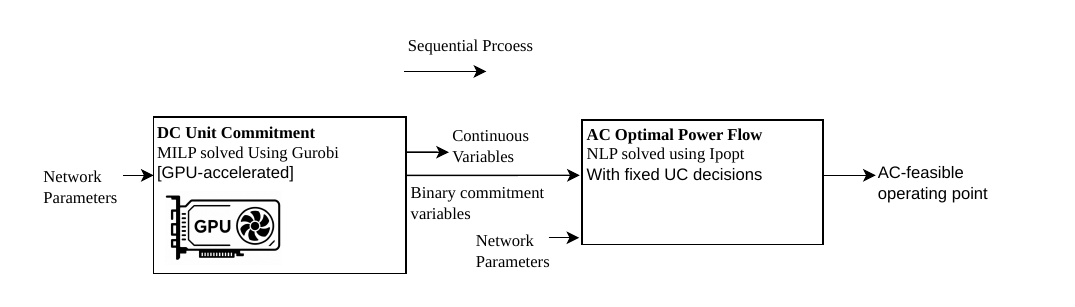}
    \caption{The two-stage framework builds on our work in \cite{TPEC, IEEE-TIA}. In this work, GPUs are used to accelerate the DC stage for large-scale networks.}
    \label{gpu-gurobi}
\end{figure*}

\subsection{Dataset}

To evaluate the proposed methodology under realistic and computationally demanding conditions, this work utilizes a repository of large-scale benchmark instances released by ARPA-E as part of the GO Competition, which is available in \cite{go_data}. The study focuses on Trial 3 test systems with network sizes of 4224, 6049, and 6717 buses. %

Each system is examined across multiple operational planning horizons defined in the ARPA-E GO Competition benchmark \cite{go_data}. The dataset is organized into three divisions, each with a distinct time horizon and temporal resolution. Division 1 corresponds to an 8-hour short-term horizon discretized into 18 time steps with mixed resolutions (0.25 h, 0.5 h, and 1 h). Division 2 corresponds to a 48-hour horizon discretized into 48 one-hour time steps. Division 3 corresponds to a 7-day (168-hour) horizon discretized into 42 four-hour time steps.

The dimensionality of the resulting optimization problems grows rapidly with both network size and horizon length. As a representative example, the  8-hour look-ahead instance of the 4224-bus system contains on the order of one million continuous variables, a comparable number of constraints, and over 100,000 binary variables. The breakdown of variable and constraint counts for all cases is provided in Table \ref{tab01}. The tabulated values reflect the baseline Division 1 configuration with 18 time steps; problem sizes for longer horizons are obtained by scaling these quantities to 48 time steps for Division 2 and 42 time steps for Division 3.
 
In total, the initial computational study comprises 57 distinct optimization instances, covering all combinations of network scale and scheduling horizon considered. 

\begin{table}[b]
\caption{Dimensions of the optimization problems for the three test networks under Division 1 (18 time steps). To obtain equivalent counts for Division 2 (48 steps), multiply by $48/18$; for Division 3 (42 steps), multiply by $42/18$ \cite{TPEC26}.}
\vspace{-.25cm}
\begin{center}
\small
\begin{tabular}{|l||c|c|c|}
\hline
\multirow{2}{*}{\textbf{Variable / Metric}} 
& \multicolumn{3}{c|}{\textbf{System}} \\
\cline{2-4}
& \textbf{4224-Bus} & \textbf{6049-Bus} & \textbf{6717-Bus} \\
\hline
\hline
\# of Binary Variables & 116,154 & 67,932 & 104,868 \\
\# of Continuous Variables & 966,220 & 1,673,301 & 2,165,998 \\
\# of Constraints & 1,168,114 & 1,580,545  & 2,003,974 \\
\hline
\end{tabular}
\label{tab01}
\end{center}
\end{table}

\subsection{Reference Model}
The reference framework developed in our prior studies \cite{TPEC, IEEE-TIA, NAPS2025} combines both DC and AC components (shown in Figure~\ref{gpu-gurobi}) and has demonstrated robust performance, with average scaled scores exceeding 0.98 across various network sizes and operational scenarios. The scaled score quantifies the closeness of the computed solution to the best-known benchmark, and maintaining values around 0.98 over approximately 300 scenarios illustrates the consistency and reliability of the model.

In the present work, we enhance the computational efficiency of the unit commitment component (the first component shown in %
 Figure~\ref{gpu-gurobi})
by integrating GPU-accelerated optimization with the Gurobi solver. This implementation significantly speeds up calculations for large-scale networks, including the 4224-, 6049-, and 6717-bus systems.

\subsection{Hardware and Computing Setup}

The implementation of the PDHG algorithm with Gurobi in a GPU-accelerated setup requires NVIDIA GPU hardware—ideally an NVIDIA H100, as recommended in the Gurobi documentation \cite{gurobi}—along with CUDA version 12.9 and a 64-bit Linux operating system. Although Gurobi supports Windows for general usage, GPU-based acceleration is currently limited to Linux platforms.

In GPU-enabled mode, the PDHG iterations are executed on the graphics card, while the CPU manages higher-level operations, including presolve routines, crossover steps, simplex computations, and branch-and-bound procedures. Leveraging GPUs for optimization can yield significant performance improvements compared to conventional CPU-only execution.

For the experiments, an NVIDIA A800 GPU with 40 GB of memory was employed, operating in a Linux64 environment through the Windows Subsystem for Linux (WSL). All optimization computations were carried out using Gurobi 13.0.0 beta 2. The computing node was built around an Intel\textregistered{} Xeon\textregistered{} W9-3495X processor featuring 56 physical cores and 112 threads, with a base frequency of 1.90 GHz and a turbo boost up to 4.80 GHz, supported by 256 GB of system memory to manage the computational demands of the datasets. Gurobi was configured to run with 64 threads, double the default 32-thread setting. Increasing the thread count beyond this did not lead to performance improvements, likely due to limitations in memory bandwidth and the overhead associated with thread synchronization.

Table~\ref{tab:hardware} provides an overview of the hardware and software configurations used in our experiments. In this setup, the GPU handled the iterative computations of the PDHG algorithm, while the CPU managed other solver components, including branch-and-bound and crossover procedures, using multithreading. The GPU and multi-core processor thus worked together to reduce overall computation time. In contrast, when alternative approaches for solving the relaxed UC, such as linear programming or barrier methods, are applied, the computations rely almost entirely on CPU-based parallelism, with the GPU providing little or no acceleration.

\begin{table}[t]
\centering
\small
\caption{System Hardware and Software Configuration Used in Experiments \cite{TPEC26}}
\label{tab:hardware}
\begin{tabular}{|l||l|}
\hline
\textbf{Component} & \textbf{Details} \\ \hline
\hline
GPU & NVIDIA A800 with 40 GB VRAM \\ 
CPU & Intel\textregistered{} Xeon\textregistered{} W9--3495X, 56 cores\\ 
 & / 112 threads, 1.90--4.80 GHz \\ 
Memory& 256 GB RAM \\ 
Operating System & 64--bit Linux (via WSL on Windows) \\ 
CUDA & Version 12.9 \\ 
Python & Version 3.11 \\ 
Gurobi Solver& Version 13.0.0 beta 2\\ 
CPU Threads-- & 64 (default: 32) \\ 
Gurobi & \\ \hline
\end{tabular}
\end{table}

\textcolor{black}{The GPU implementation primarily accelerates the computationally intensive operations within the PDHG iterations. In each iteration, the primal and dual updates involve sparse matrix-vector products and element-wise vector operations, which are well suited to massively parallel GPU execution. The problem matrices and associated vectors are maintained in GPU memory during the iterative solution process to minimize repeated CPU--GPU data transfers. Thus, the computational workload associated with the PDHG iterations is executed on the GPU, while the CPU remains responsible for solver-level operations and coordination.}

\section{Results and Discussion}

\subsection{Analysis of Solver Runtime}

\begin{figure}[!b]
    \centering
    \includegraphics[width=0.98\linewidth]{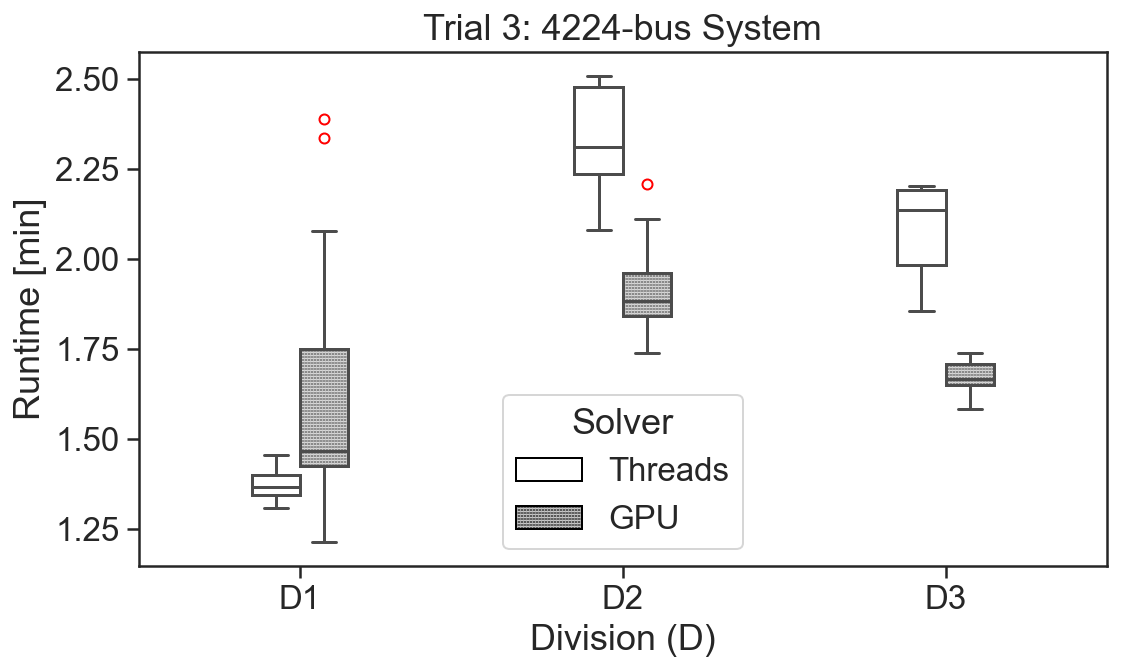}
    \caption{Runtime distribution for the 4224-bus system from Trial 3 (Divisions 1–3). The GPU-enabled solver demonstrates faster convergence and smaller fluctuations for Divisions 2 and 3. `Threads' denotes the barrier method executed on the CPU using multi-threading, while `GPU' denotes the PDHG method executed on the GPU. Outlier observations are represented by hollow red circular markers.
}
    \label{fig:4224}
\end{figure}

\begin{figure}[!b]
    \centering
    \includegraphics[width=0.9\linewidth]{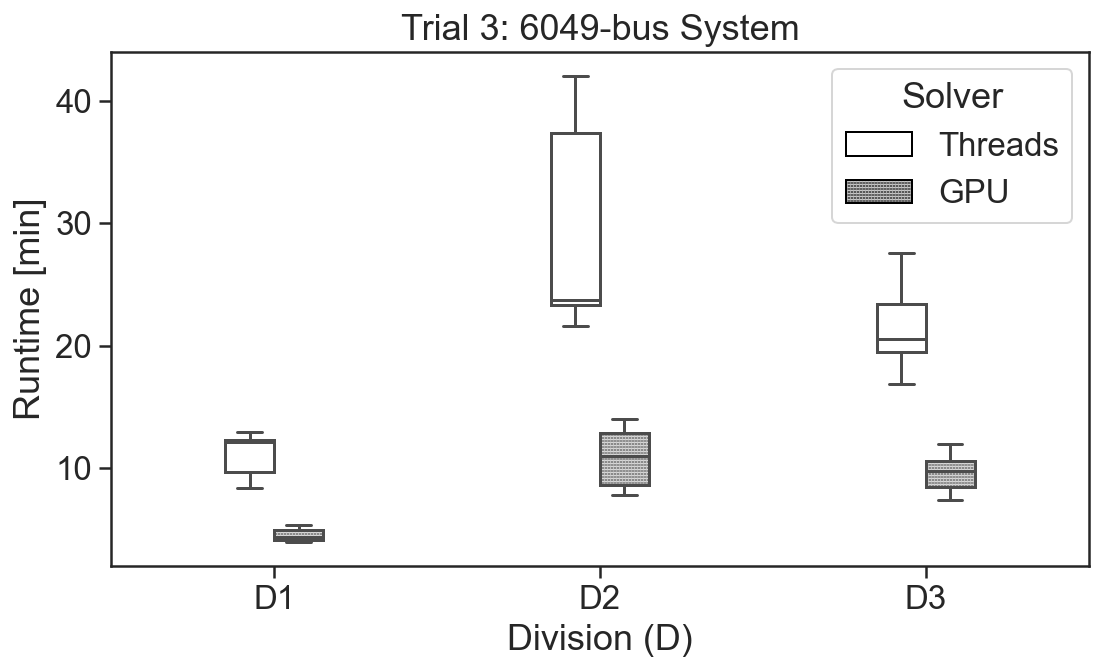}
    \caption{Runtime distribution for the 6049-bus system from Trial 3 (Divisions 1–3). The GPU-enabled solver demonstrates faster convergence and smaller fluctuations across all Divisions.
}
    \label{fig:6049}
\end{figure}

\begin{figure}
    \centering
    \includegraphics[width=0.98\linewidth]{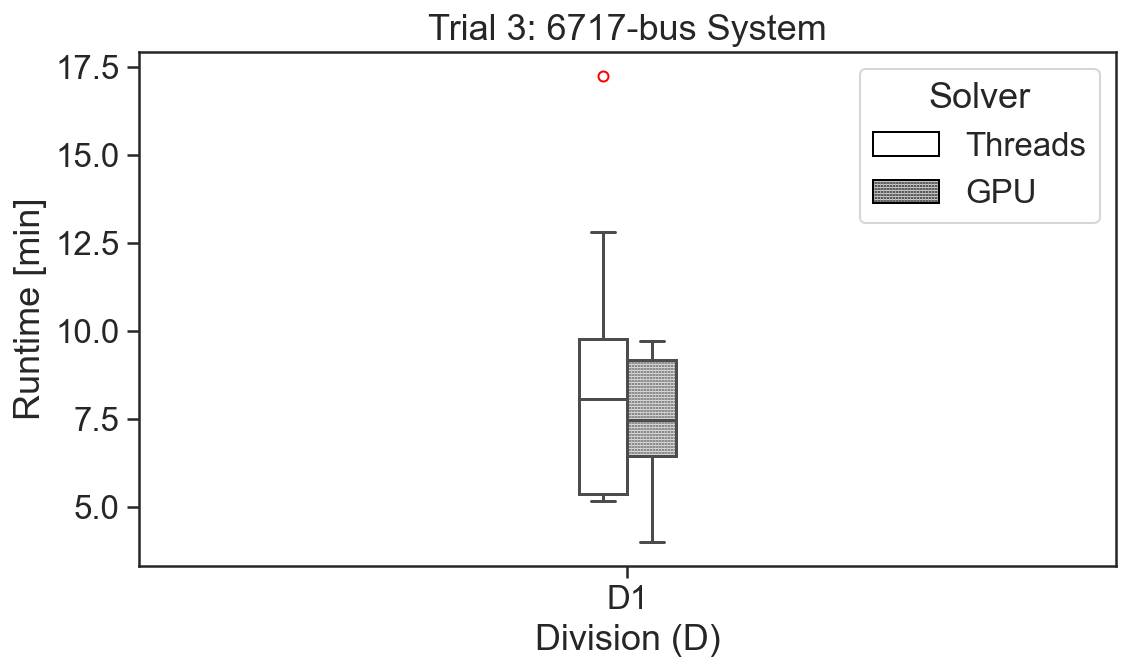}
    \caption{Runtime distribution for the 6717-bus system from Trial 3 (Division 1). Overall, the GPU-accelerated solver exhibits faster convergence.
}
    \label{fig:6717}
\end{figure}

Solver runtimes were assessed for the 4224-, 6049-, and 6717-bus test systems over multiple scenarios, across Divisions 1–3, and for different solution strategies. The comparison focuses on a GPU-accelerated implementation based on PDHG and a multi-threaded CPU implementation relying on a barrier algorithm. The distributions of the solution times for each network and division are summarized using boxplots in Figures \ref{fig:4224}, \ref{fig:6049}, and \ref{fig:6717}. In general, the GPU-based approach yields shorter execution times, with the most pronounced benefits observed for larger networks and for divisions associated with higher computational complexity, such as those involving extended control horizons in Divisions 2 and 3.

For the 4224-bus case, the PDHG-based solver running on the GPU consistently outperforms the CPU-based solver in Divisions 2 and 3. The performance gap becomes more evident in the 6049-bus system, where the CPU solver exhibits runtime spikes for certain scenarios, while the GPU-based solver maintains lower solution times with reduced variance across all divisions and scenarios. In the 6717-bus system, GPU implementation demonstrates superior overall efficiency relative to the multi-threaded CPU solver.

The observed runtime reductions primarily stem from the improved initial solutions generated by PDHG, which serve as effective warm starts for the subsequent crossover phase, thereby reducing the number of iterations required for convergence. Although the barrier-based CPU solver leverages parallelism through multi-threading, it remains limited by CPU-centric execution and cannot fully utilize the fine-grained parallelism available on GPUs. Variability across scenarios is evident in the boxplots, highlighting differences in convergence behavior between the two solution strategies. Overall, the GPU-based PDHG method offers the largest performance gains for large-scale, computation-intensive cases—most notably for Divisions 2 and 3 of the 6049-bus system—delivering faster runtimes with improved consistency.

\vspace{-0.2cm}

\subsection{PDHG on GPU vs. CPU}
\label{gpuvscpu}
\begin{figure}
    \centering
    \includegraphics[width=0.98\linewidth]{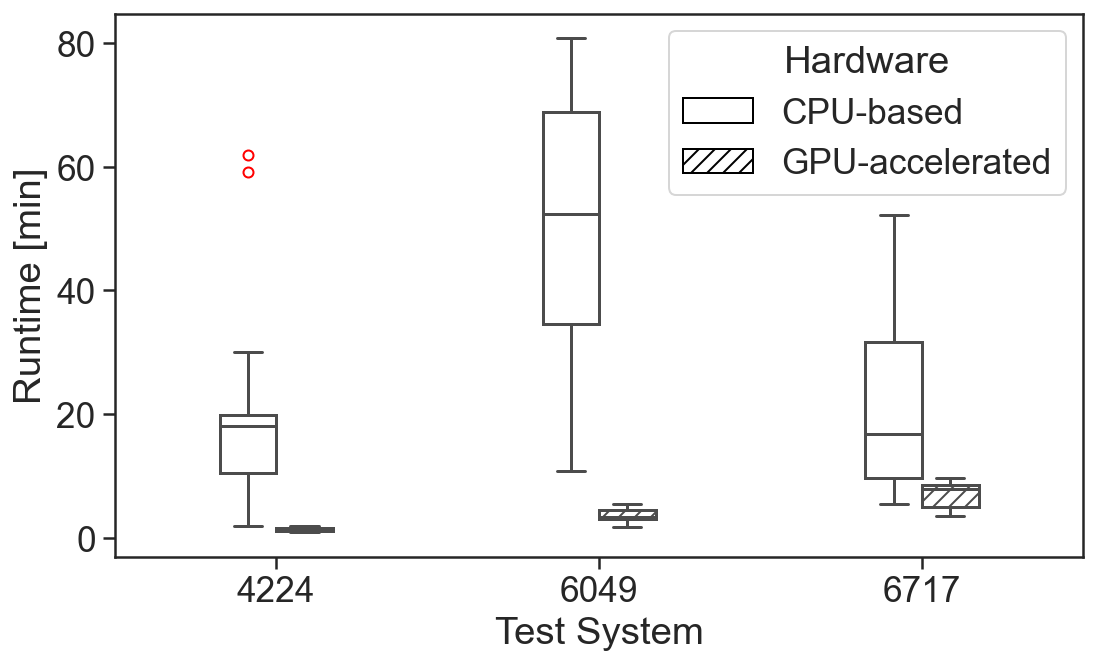}
    \caption{\textcolor{black}{Runtime distribution comparison of CPU-based and GPU-accelerated PDHG implementations for division 1.%
    }}
    \label{GPU_CPU}
\end{figure}

\textcolor{black}{
Although the barrier method achieves better performance than PDHG when both are executed on the CPU, its computational operations are not well suited for GPU architectures due to limited parallelism and algorithmic characteristics. In contrast, PDHG relies on matrix-vector operations and element-wise updates that can be efficiently parallelized and fully executed on GPUs. To isolate the effect of GPU acceleration from the choice of optimization algorithm, the performance of the same PDHG solver implemented on CPU and GPU is compared for Division 1. The runtime comparison is presented in Figure~\ref{GPU_CPU} and Tables~\ref{tab:pdhg_avg_runtime} and~\ref{tab:pdhg_max_runtime}. The GPU-based PDHG implementation significantly reduces computational time, achieving average runtime speedups ranging from 2.9$\times$ to 14.6$\times$ compared with the CPU-based PDHG implementation. The maximum runtime comparison demonstrates even greater improvements, with speedups reaching up to 30.24$\times$. %
}

\begin{table}[!]
\caption{\textcolor{black}{Average runtime comparison between CPU- and GPU-based PDHG solver for division 1. 
Colors indicate GPU performance improvement (blue). Runtime values are reported in minutes.}}
\label{tab:pdhg_avg_runtime}
\centering
\begin{tabular}{lllll}
\hline
\textbf{Bus} & 
\textbf{PDHG (CPU)} & 
\textbf{PDHG (GPU)} & 
\textbf{Speedup} \\
\hline
4224 & 20.7 & 1.4 & \textcolor{blue}{14.6\texttt{x}} \\
6049 & 50.0 & 3.7 & \textcolor{blue}{13.4\texttt{x}} \\
6717 & 20.5 & 7.0 & \textcolor{blue}{2.9\texttt{x}} \\
\hline
\end{tabular}
\end{table}

\begin{table}[!]
\caption{\textcolor{black}{Maximum Runtime Comparison Between CPU- and GPU-Based PDHG Solver for Division 1. 
Arrows indicate GPU performance improvement ($\downarrow$, blue) compared with the CPU implementation. 
Runtime values are reported in minutes.}}
\label{tab:pdhg_max_runtime}
\centering
\begin{tabular}{llll}
\hline
\textbf{Bus} & 
\textbf{PDHG (CPU)} & 
\textbf{PDHG (GPU)} \\
\hline
\hline
4224 & 61.9 & 2.0 \textcolor{blue}{$\downarrow$} \\
6049 & 80.7 & 5.6 \textcolor{blue}{$\downarrow$} \\
6717 & 52.2 & 9.7 \textcolor{blue}{$\downarrow$} \\
\hline
\end{tabular}
\end{table}

\subsection{PDHG Execution \& Crossover Refinement}

In this study, default stopping tolerances were used for PDHG (see the termination criteria described in \cite{gurobi}). After these conditions are met, a crossover procedure is applied to convert the PDHG output into a basic feasible solution that can be used reliably in later optimization stages.

Unlike the simplex method, which naturally maintains a basic feasible solution, PDHG and interior-type methods produce dense solutions that may contain small feasibility or optimality errors. Instead of switching to the simplex algorithm—which is typically slower and poorly suited for GPU execution—the crossover step efficiently transforms the PDHG result into a sparse and well-conditioned basic solution. Only the PDHG iterations are executed on the GPU; all crossover operations are performed on the CPU.

The choice of PDHG stopping tolerances affects overall performance. Using looser tolerances, such as a larger allowable duality gap, reduces the time spent in PDHG iterations. However, this leads to less accurate intermediate solutions, requiring the crossover step to spend more time correcting feasibility and optimality. This trade-off must be considered when selecting termination settings.

First-order optimization algorithms, such as PDHG, exhibit linear convergence and are therefore not well suited for achieving very tight absolute accuracy. In practice, PDHG is terminated once relative feasibility metrics reach acceptable levels. To obtain a solution that meets strict accuracy requirements, Gurobi automatically applies a crossover procedure that systematically refines the PDHG output into a highly accurate and feasible solution.
Although the crossover phase is executed on the CPU, this stage benefits significantly from the improved solution quality produced by the GPU-accelerated PDHG phase. Specifically, PDHG delivers a strong warm start for the crossover step and generates a higher-quality initial basis. As a result, the remaining CPU-based \textcolor{black}{crossover }computations complete more efficiently, since they begin from a better-conditioned and more accurate solution.
\begin{figure*}[ht]
    \centering
    \begin{subfigure}[b]{0.48\linewidth}
        \centering
        \includegraphics[width=0.87\linewidth]{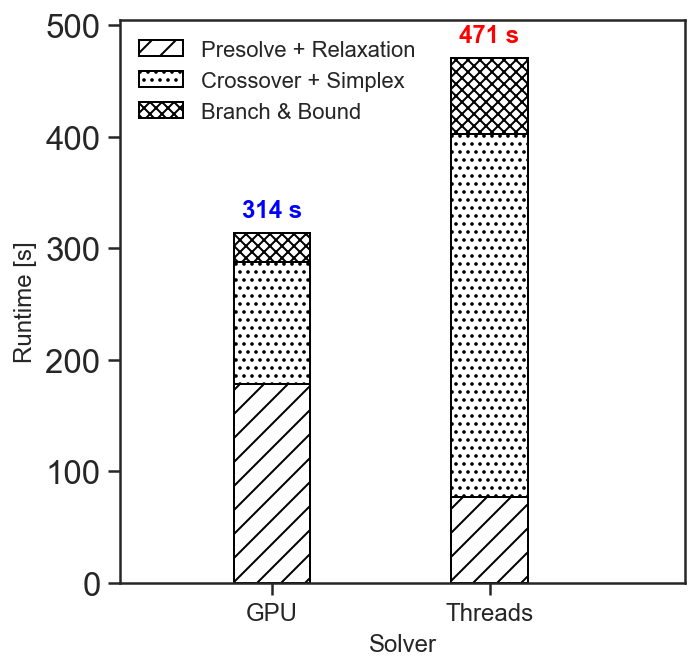}
        \caption{Random Scenario 1}
        \label{fig:Time_breakdown1}
    \end{subfigure}
    \hfill
    \begin{subfigure}[b]{0.48\linewidth}
        \centering
        \includegraphics[width=0.87\linewidth]{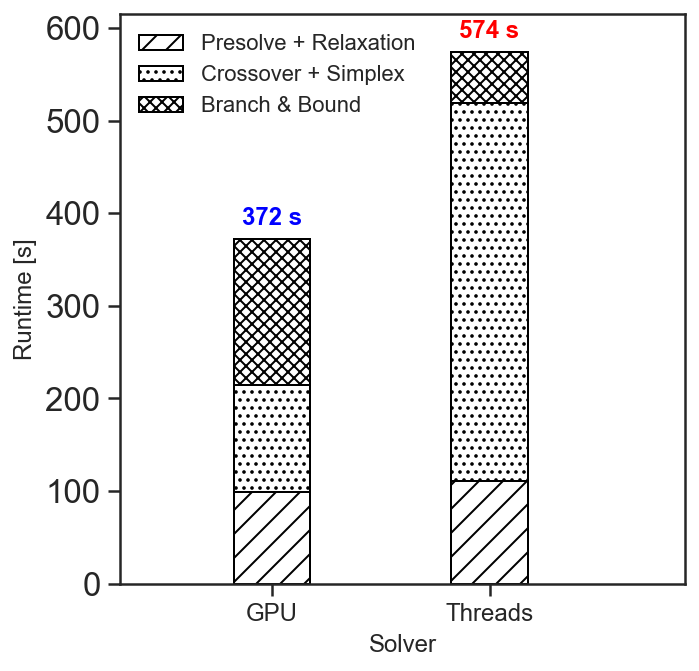}
        \caption{Random Scenario 2}
        \label{fig:Time_breakdown2}
    \end{subfigure}
    \caption{Runtime comparison between the GPU-accelerated and multi-threaded solvers for the 6049-bus system in Trial 3 under two randomly selected scenarios. Each bar shows the total solution time, partitioned into three components: (i) presolve and problem relaxation, (ii) crossover and simplex iterations, and (iii) branch-and-bound. In the Threads-based implementation, the relaxation phase is handled using a barrier method, whereas the GPU-based solver employs PDHG. }
    \label{fig:Time_breakdown_combined}
\end{figure*}

Figure~\ref{fig:Time_breakdown_combined} shows the breakdown of computation times for the GPU-based solver and the Threads-based solver for two randomly selected 6049-bus scenarios. The GPU-based solver achieves a shorter total runtime, with the largest improvements occurring in the crossover phase, which takes the most time on the CPU.

In the first scenario, the CPU-based solver completes the presolve and relaxation stages (using the barrier algorithm) faster than the GPU. However, the GPU-based PDHG relaxation produces a higher-quality intermediate solution, which greatly reduces the time needed for the crossover  phase, lowering the total runtime. In the second scenario, the GPU completes the relaxation stage slightly faster. Although branch-and-bound takes a bit longer, the large reduction in crossover time still gives the GPU solver an overall speed advantage.

These gains result from the higher-quality solutions produced by PDHG, which offer a more effective warm start and a better initial basis for the crossover procedure. In particular, the crossover phase initialized from the GPU-based relaxation runs over $2.96\times$ faster than when starting from the CPU-based relaxation. Overall, across the two scenarios, the GPU solver achieves approximately a $1.5\times$ speedup compared to the CPU solver.

\subsection{Summary of Achieved Speedups}
The runtime results demonstrate the computational advantages of leveraging GPU acceleration for solving the relaxed mixed-integer linear program associated with the unit commitment problem. Tables~\ref{tab:max_times} and~\ref{tab:avg_speedup} report the maximum runtimes and corresponding acceleration factors obtained using the GPU-based approach. Across the evaluated test cases, the GPU implementation generally lowers both worst-case and average runtimes, with the exception of Division 1 for the 4224-bus system. Specifically, reductions are observed for the 4224-bus system in Divisions 2 and 3, where speedups exceed $1.2\times$, for the 6049-bus system across all divisions with speedups greater than $2.27\times$, and for the 6717-bus system with a speedup of $1.23\times$. A particularly notable improvement occurs for the 6049-bus network in Division 2, where the maximum runtime is reduced from about 42 minutes to 14 minutes.

\subsection{Assessment of Solution Quality}

The numerical results indicate that the GPU-accelerated solver maintains solution quality, as summarized in Table~\ref{tab:obj_score_distribution1}. \textcolor{black}{The feasibility of the GPU-generated solutions is also preserved, with no constraint violations observed across the 57 evaluated scenarios.} The table reports the distribution of objective scores, computed as the objective value obtained with the GPU-based solver divided by that of the multi-threaded CPU solver. Because the objective function represents the maximization of market surplus \cite{GO3}, score values exceeding unity correspond to superior solutions.

Among the 57 evaluated scenarios \textcolor{black}{from Trial 3}, 36 scenarios yield objective scores of at least 1. The remaining 21 scenarios produce scores marginally below 1, yet consistently above 0.99985. The average objective score across all scenarios is approximately 0.999986, confirming that the GPU-based approach achieves objective values that are essentially equivalent to those of the multi-threaded CPU solver.

\textcolor{black}{The PDHG solver running on the GPU produces objective values that are nearly identical to those obtained by the CPU-based PDHG solver from Section~\ref{gpuvscpu}. Among the 42 tested scenarios, 36 scenarios have an objective-value ratio equal to 1, three scenarios have ratios slightly greater than 1, and only three scenarios have ratios below 1, while remaining above 0.99995. The small differences in objective values can be attributed to numerical precision and differences in floating-point arithmetic between the CPU and GPU implementations, including differences in the order of parallel floating-point operations.
}

 \subsection{Reducing Spin/Wait Overhead \\in Multi-Threaded CPU Solvers}
\begin{table}
\caption{\textcolor{black}{Maximum runtimes for the Threads- and GPU-based solvers \cite{TPEC26}. Arrows show whether the GPU solver is faster ($\downarrow$, blue) or slower ($\uparrow$, red) than the Threads implementation.}}
\centering
\small
\setlength{\tabcolsep}{4pt}
\begin{tabular}{|l||l|c|c|}
\hline
\textbf{Network} & \textbf{Division (D)} & \multicolumn{2}{c|}{\textbf{Maximum Runtime [min]}}  \\
\cline{3-4}
 & & Threads & GPU  \\
\hline
\hline
4224-bus & D1 & 1.46 & 2.39 \textcolor{red}{$\uparrow$} \\
     & D2 & 2.51 & 2.21 \textcolor{blue}{$\downarrow$} \\
     & D3 & 2.20 & 1.74 \textcolor{blue}{$\downarrow$} \\
\hline
6049-bus & D1 & 12.94 & 5.41 \textcolor{blue}{$\downarrow$} \\
     & D2 & 42.03 & 14.0 \textcolor{blue}{$\downarrow$} \\
     & D3 & 27.57 & 12.0 \textcolor{blue}{$\downarrow$} \\
\hline
6717-bus & D1 & 17.24 & 9.71 \textcolor{blue}{$\downarrow$} \\
\hline
\end{tabular}
\label{tab:max_times}
\end{table}

In CPU-based experiments, Gurobi can execute the primal simplex, dual simplex, and barrier algorithms concurrently using multiple threads. However, when one algorithm reaches convergence, the solver may continue to wait for the remaining threads to complete, resulting in spin/wait overhead and underutilized CPU resources. This behavior was particularly evident in the 6049-bus system, as indicated in Table~\ref{tab:max_times}. To address this, a solver option was enabled to terminate the remaining algorithms as soon as a solution is found, thereby minimizing idle CPU cycles and providing a more accurate measure of computational performance. This modification improved the Threads-based runtimes and ensured a more consistent and fair comparison with the GPU-based solver. The resulting speedups of the GPU solver relative to the adjusted CPU implementation are reported in Table~\ref{tab:avg_speedup_updated}, with average acceleration factors ranging from 1.31× to 1.93×. Even after accounting for the spin/wait adjustment, the GPU solver maintains superior performance.

\begin{table}
\caption{\textcolor{black}{Average runtimes for Threads versus GPU solvers and the achieved speed-up \cite{TPEC26}. Blue indicates cases where GPU is faster, red where it is slower. }}
\centering
\small
\setlength{\tabcolsep}{4pt}
\begin{tabular}{|l||l|c|c|c|}
\hline
\textbf{Network} & \textbf{Division (D)} & \multicolumn{2}{c|}{\textbf{Avg. Runtime [min]}} & \textbf{Avg.} \\
\cline{3-4}
 & & Threads & GPU&\textbf{Speed-up} \\
\hline
\hline
4224-bus &  D1 & 1.37 & 1.61 & \textcolor{red}{$0.85\texttt{x}$} \\
     & D2 & 2.34 & 1.92 & \textcolor{blue}{$1.22\texttt{x}$} \\
     & D3 & 2.08 & 1.67 & \textcolor{blue}{$1.25\texttt{x}$} \\
\hline
6049-bus & D1 & 11.13 & 4.56 & \textcolor{blue}{$2.44\texttt{x}$} \\
     & D2 & 29.36 & 10.19 & \textcolor{blue}{$2.88\texttt{x}$} \\
     & D3 & 21.85 & 9.64 & \textcolor{blue}{$2.27\texttt{x}$} \\
\hline
6717-bus & D1 & 8.76 & 7.12 & \textcolor{blue}{$1.23\texttt{x}$} \\
\hline
\end{tabular}
\label{tab:avg_speedup}
\end{table}

The impact of terminating idle algorithms on average CPU runtimes is summarized in Table~\ref{tab:CPU_avg_speedup}, expressed as the resulting speedups, demonstrating more efficient utilization of multithreaded resources. These findings underscore the importance of managing concurrent solver behavior to optimize CPU performance and enable a fair comparison with GPU-accelerated implementations.

\begin{table}[b]
\caption{\textcolor{black}{Distribution of GPU objective scores, defined as the ratio between GPU and Threads objective values \cite{TPEC26}. The scores demonstrate that the GPU-based solver achieves objective values essentially equivalent to those of the CPU-based solver.}}
\centering
\small
\begin{tabular}{|l||c|}
\hline
\textbf{Score Range} & \textbf{Scenario Count} \\
\hline
\hline
\textbf{Score $\geq 1$} & \textcolor{blue}{36} \\
$0.99999 <$ Score $< 1$ & \textcolor{red}{16} \\
$0.99985 <$ Score $< 0.99999$ & \textcolor{red}{5} \\
\hline
\hline
\textbf{Mean Score} & $\approx$ \textcolor{red}{$0.999986$} \\
\hline
\end{tabular}
\label{tab:obj_score_distribution1}
\end{table}

\begin{table}[h!]
\caption{\textcolor{black}{Average runtimes for improved Threads solver versus GPU solver and the achieved speed-up, for the 6049-bus system \cite{TPEC26}. The runtime of the Threads solver was reduced by mitigating CPU spin and wait overhead.}}
\centering
\small
\setlength{\tabcolsep}{4pt}
\begin{tabular}{|l||l|c|c|c|}
\hline
\textbf{Division (D)} & \multicolumn{2}{c|}{\textbf{Avg. Runtime [min]}} & \textbf{Avg.} \\
\cline{2-3}
 & Threads & GPU&\textbf{Speed-up} \\
\hline
\hline
 D1 & 5.97 & 4.56 & \textcolor{blue}{$1.31\texttt{x}$} \\
 D2 & 19.7 & 10.19 & \textcolor{blue}{$1.93\texttt{x}$} \\
D3 & 16.06 & 9.64 & \textcolor{blue}{$1.67\texttt{x}$} \\
\hline
\end{tabular}
\label{tab:avg_speedup_updated}
\end{table}

\begin{table}[h!]
\caption{\textcolor{black}{Runtime improvements for the Threads-based solver obtained by mitigating CPU spin–wait overhead in the 6049-bus system \cite{TPEC26}.}}
\centering
\small
\setlength{\tabcolsep}{4pt}
\begin{tabular}{|l||c|}
\hline
\textbf{Division (D)} & \textbf{Avg. Speed-up} \\
\hline
\hline
 D1  & \textcolor{blue}{$1.86\texttt{x}$} \\
 D2  & \textcolor{blue}{$1.49\texttt{x}$} \\
D3 & \textcolor{blue}{$1.36\texttt{x}$} \\
\hline
\end{tabular}
\vspace{-.3cm}
\label{tab:CPU_avg_speedup}
\end{table}

\subsection{Validation Using Trial 2 Dataset}
The Trial 2 dataset contains 59 scenarios for the 4224-, 6049-, and 6717-bus systems across Divisions 1–3. The GPU-based solver exhibits behavior similar to that observed in Trial 3 in terms of runtime performance and solution quality. The detailed results for Trial 2 are further discussed in Section~\ref{secG}.

In the Trial 3 dataset, the code was able to solve all scenarios. The 6049-bus network from the Trial 2 dataset is more challenging to solve, possibly due to ill-conditioning. In Division 2, the Threads solver fails on scenarios 23 and 32, whereas the GPU solver successfully solves them, highlighting how introducing a GPU solver can increase the number of solvable scenarios. Both solvers fail on scenarios 31 and 33. In Division 3, the Threads solver fails on scenario 21, while the GPU solver fails on different scenarios (32 and 33) that the Threads solver successfully solves. This demonstrates the benefit of using both solvers in combination. These cases are summarized in Table~\ref{T2_fail}.

\begin{table}[b]
\centering
\small
\caption{Unsolved scenario indices for each division and solver for the 6049-bus network (Trial 2 dataset) within a 120-minute user-defined limit. Scenarios highlighted in red indicate cases where both solvers fail to converge within this time limit.}
\begin{tabular}{|c||c|l|}
\hline
\textbf{Division (D)} & \textbf{Solver} & \textbf{Unsolved Scenario Indices} \\ \hline
\hline
D2 & Threads& 23, \textcolor{red}{31}, 32, \textcolor{red}{33} \\ 
D2 & GPU& \textcolor{red}{31}, \textcolor{red}{33}       \\ 
\hline
D3 & Threads& 21          \\ 
D3 & GPU& 32, 33       \\ \hline
\end{tabular}
\label{T2_fail}
\end{table}

\subsection{Concurrent GPU--Threads Solver}
\label{secG}

\begin{figure*}
    \centering
    \resizebox{\solverfigwidth}{!}{%

\begin{tikzpicture}[
  x=1cm,y=1cm,
  font=\sffamily,
  >=Latex,
  box/.style={draw=paperline,rounded corners=3pt,thick,align=center,inner xsep=8pt,inner ysep=6pt,fill=white,text=black},
  topbox/.style={box, fill=papergrayfill},
  smallbox/.style={draw=paperline,rounded corners=3pt,thick,align=center,inner xsep=7pt,inner ysep=5pt,fill=white},
  monitorbox/.style={smallbox, draw=paperpurple, fill=paperpurplefill},
  stage/.style={draw=paperline,rounded corners=2pt,thick,align=center,inner xsep=6pt,inner ysep=4pt,fill=white,minimum width=3.9cm,minimum height=0.62cm},
  stagecommon/.style={stage, fill=papergrayfill},
  stagegpu/.style={stage, draw=paperblue, fill=paperbluefill},
  stagebarrier/.style={stage, draw=papergreen, fill=papergreenfill},
  decision/.style={draw=paperpurple,diamond,aspect=1.8,thick,align=center,inner xsep=3pt,inner ysep=2pt,fill=paperpurplefill},
  main/.style={-Latex,thick,draw=paperline},
  monitor/.style={-Latex,thick,dashed,draw=paperpurple},
  solverframe1/.style={draw=paperblue, rounded corners=4pt, thick},
  solverframe2/.style={draw=papergreen, rounded corners=4pt, thick},
  labelpill1/.style={draw=paperblue, rounded corners=8pt, fill=paperbluefill, inner xsep=6pt, inner ysep=4pt},
  labelpill2/.style={draw=papergreen, rounded corners=8pt, fill=papergreenfill, inner xsep=6pt, inner ysep=4pt}
]

\node[font=\bfseries\Large, text=paperline] at (0,7.25) {Concurrent CPU--GPU Solver Framework};

\node[topbox, minimum width=10.0cm] (cores) at (0,6.0)
{Partition 48 CPU cores evenly between two solver instances};

\node[topbox, minimum width=9.5cm] (launch) at (0,4.70)
{Launch both solver instances on the same optimization problem\\with an identical optimality-gap tolerance};

\coordinate (fork) at (0,3.70);
\draw[main] (cores.south) -- (launch.north);
\draw[thick,draw=paperline] (launch.south) -- (fork);
\fill[paperline] (fork) circle (1.2pt);

\node[font=\bfseries, text=paperpurple] at (0,3.20) {Periodic supervisory loop};
\node[monitorbox, minimum width=4.5cm] (monitorstart) at (0,2.35) {Start periodic status check};
\node[monitorbox, minimum width=2.7cm] (wait) at (0,0.90) {Wait 0.5 s};
\node[decision, minimum width=2.05cm, minimum height=1.00cm] (check) at (0,-0.95) {Has either solver\\converged?};
\node[monitorbox, minimum width=5.35cm] (terminate) at (0,-4.10)
{Terminate the other solver and return the converged solution};

\draw[main] (fork) -- (monitorstart.north);
\draw[main] (monitorstart.south) -- ++(0,-0.12) -- (wait.north);
\draw[main] (wait.south) -- ++(0,-0.22) -- ($(check.north)+(0,0.14)$) -- (check.north);
\draw[main] (check.south) -- (terminate.north);
\path ($(check.south)!0.52!(terminate.north)+(0.36,0)$) node[font=\small, text=paperpurple] {Yes};

\coordinate (loopLeft) at ($(check.west)+(-1.00,0)$);
\coordinate (loopUp) at ($(loopLeft)+(0,1.85)$);
\draw[main] (check.west) -- (loopLeft) -- (loopUp) -- (wait.west);
\path ($(check.west)!0.55!(loopLeft)+(0,0.18)$) node[font=\small, text=paperpurple] {No};

\node[labelpill1, font=\bfseries\large, text=paperblue] (label1) at (-6.4,2.54) {Solver 1};
\node[labelpill2, font=\bfseries\large, text=papergreen] (label2) at ( 6.4,2.54) {Solver 2};

\node[stagecommon] (s11) at (-6.4,1.32) {Presolve};
\node[stagegpu, below=0.34cm of s11] (s12) {PDHG (GPU)};
\node[stagecommon, below=0.34cm of s12] (s13) {Crossover};
\node[stagecommon, below=0.34cm of s13] (s14) {Simplex (if needed)};
\node[stagecommon, below=0.34cm of s14] (s15) {Branch-and-Bound};

\draw[main] (s11.south) -- (s12.north);
\draw[main] (s12.south) -- (s13.north);
\draw[main] (s13.south) -- (s14.north);
\draw[main] (s14.south) -- (s15.north);

\node[solverframe1, fit=(s11)(s12)(s13)(s14)(s15), inner sep=12pt] (solver1box) {};

\node[stagecommon] (s21) at (6.4,1.32) {Presolve};
\node[stagebarrier, below=0.34cm of s21] (s22) {Barrier};
\node[stagecommon, below=0.34cm of s22] (s23) {Crossover};
\node[stagecommon, below=0.34cm of s23] (s24) {Simplex (if needed)};
\node[stagecommon, below=0.34cm of s24] (s25) {Branch-and-Bound};

\draw[main] (s21.south) -- (s22.north);
\draw[main] (s22.south) -- (s23.north);
\draw[main] (s23.south) -- (s24.north);
\draw[main] (s24.south) -- (s25.north);

\node[solverframe2, fit=(s21)(s22)(s23)(s24)(s25), inner sep=12pt] (solver2box) {};

\coordinate (s1labeltip) at (label1.north);
\coordinate (s2labeltip) at (label2.north);
\coordinate (s1elbow) at (-6.4,3.70);
\coordinate (s2elbow) at ( 6.4,3.70);
\draw[main] (fork) -- (s1elbow) -- (s1labeltip);
\draw[main] (fork) -- (s2elbow) -- (s2labeltip);

\draw[monitor] (monitorstart.west) to[out=185,in=18]  ($(solver1box.east)+(0,1.32)$);
\draw[monitor] (monitorstart.east) to[out=-5,in=162] ($(solver2box.west)+(0,1.32)$);

\end{tikzpicture}
    }
    \caption{Framework of the concurrent GPU--CPU solver, where CPU-based multi-threading and GPU-based computation are executed in parallel.}
    \label{fig:GPU-CPU}
\end{figure*}
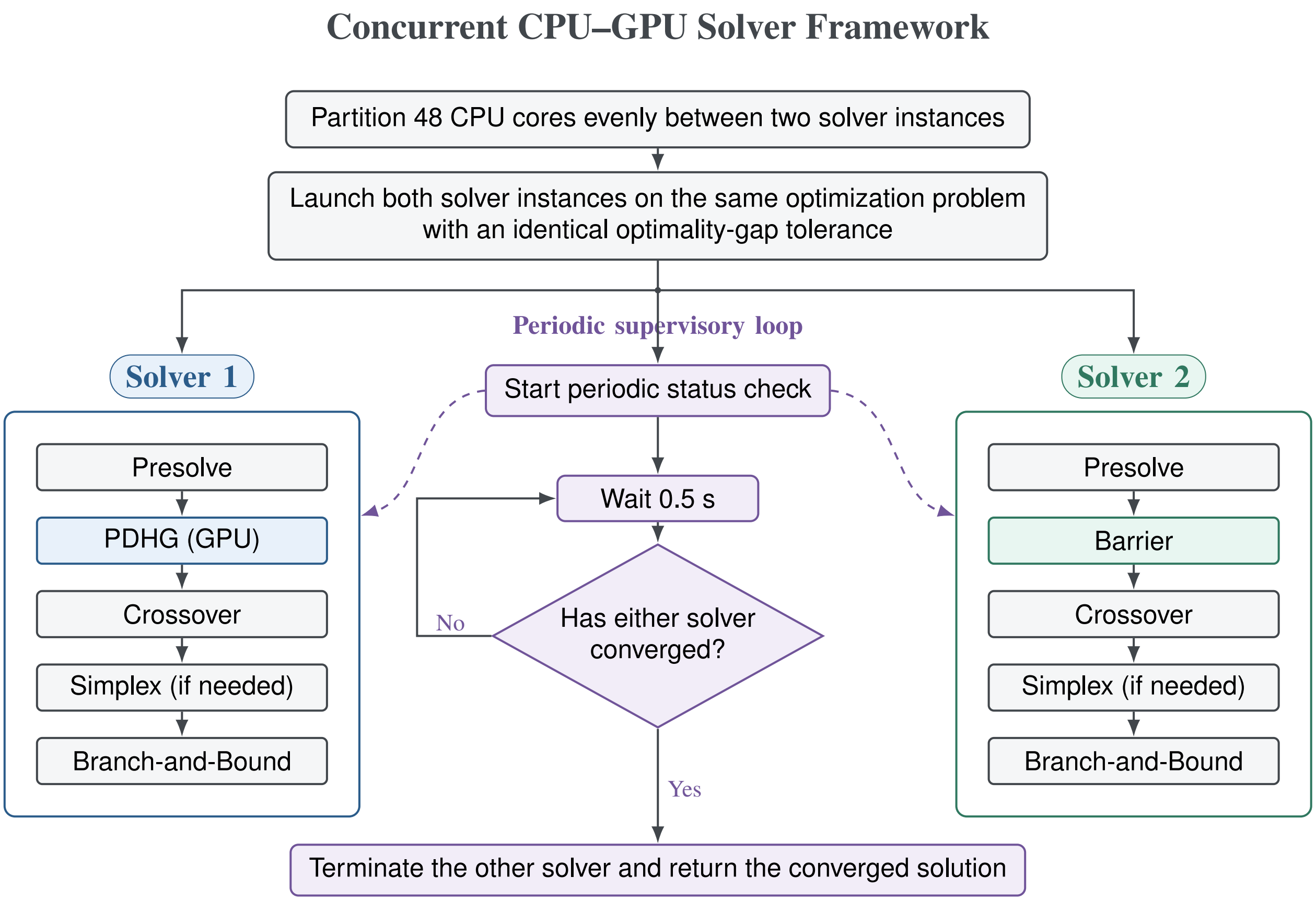

To accelerate the solution process, we ran the Threads and GPU solvers in parallel, terminating the other solver once one converged. For these experiments, we allocated 24 threads to the Threads solver and an additional 24 threads for high-level management of the GPU solver (See the framework in Figure~\ref{fig:GPU-CPU}). The threads were carefully assigned using the \texttt{$2i$} scheme, which ensures that each thread is bound to a separate physical core and avoids simultaneous use of hyperthread siblings, reducing contention and improving CPU utilization. Interestingly, reducing the number of threads from 64 (out of 112 available) to 24 for each solver did not reduce performance; in fact, in most scenarios, both solvers converged faster, likely due to reduced synchronization overhead and more efficient use of physical cores.
A small monitoring delay (0.5 seconds) was used to check which solver finishes first, without affecting solver runtimes. Once one solver converged, the other was terminated to reduce overall computation time.

Tables~\ref{T2_sync} and~\ref{T3_sync} summarize the results for the Trial 2 and Trial 3 datasets, respectively. The tables indicate the number of scenarios each solver ``won", i.e., the solver that converged first for each scenario.

\begin{table}[t]
\centering
\small
\caption{Trial 2 dataset: number of scenarios solved first by each solver in GPU/Threads concurrent mode (divisions combined).}
\begin{tabular}{|l||c|c|c|c|}
\hline
\textbf{Network} & \textbf{GPU} & \textbf{Threads} & \textbf{None} & \textbf{Total} \\ \hline
\hline
4224 & 1 & 10 & 0 & 11 \\ 
6049 & 18 & 8 & 1 & 27 \\ 
6717 & 10 & 11 & 0 & 21 \\ 
\hline
\hline
\textbf{TOTAL} & 29 & 29 & 1 & 59 \\ \hline
\end{tabular}
\label{T2_sync}
\end{table}

\begin{table}[t]
\centering
\small
\caption{Trial 3 dataset: number of scenarios solved first by each solver in GPU/Threads concurrent mode (divisions combined).}
\begin{tabular}{|l||c|c|c|}
\hline
\textbf{Network} & \textbf{GPU} & \textbf{Threads} & \textbf{Total} \\ \hline
\hline
4224 & 2 & 28 & 30 \\ 
6049 & 15 & 3  & 18 \\ 
6717 & 4 & 5  & 9  \\ 
\hline
\hline
\textbf{TOTAL} & 21 & 36 & 57 \\ \hline
\end{tabular}
\label{T3_sync}
\end{table}

These results show that, for the Trial 2 dataset, GPU and Threads solvers were roughly balanced (29 scenarios each), with only one scenario not solved by either. For Trial 3, the Threads solver dominates in most datasets, winning 36 out of 57 scenarios, while the GPU solver converged first in 21 scenarios. Using this concurrent strategy allows the solvers to complement each other, with the first to converge terminating the other, thereby reducing overall computation time.

It is important to note that the primary benefit of this concurrent strategy is the reduction in overall computation time rather than the number of scenarios each solver ``wins". For example, in the 4224-bus system, the Threads solver finishes first most of the time, but both solvers can solve all scenarios in under two minutes. In contrast, the GPU solver provides significant time savings for the 6049-bus system. By allowing the first solver to converge to terminate the other, this approach leverages the strengths of both solvers to minimize total runtime.

\begin{figure}[t]
    \centering
    \includegraphics[width=0.98\linewidth]{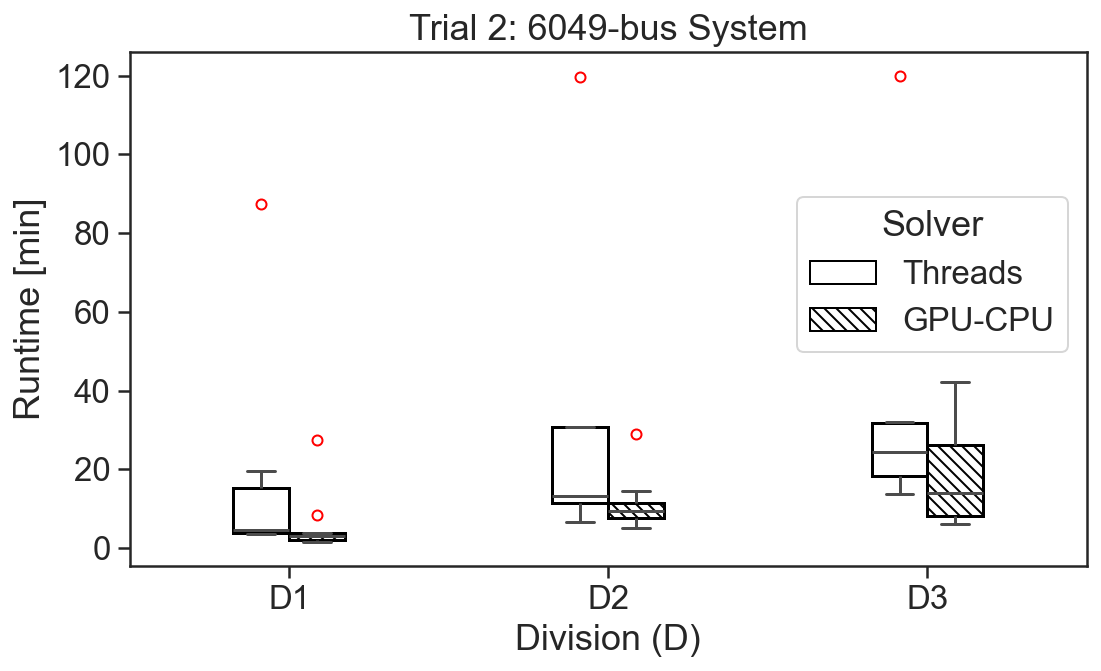}
    \caption{Runtime distribution for the 6049-bus system from Trial 2 (Divisions 1–3). The concurrent-based solver achieves faster convergence overall. `Threads' denotes the barrier method executed on the CPU using multi-threading, while `GPU-CPU' denotes the parallel execution of the CPU-based solver and the GPU-based solver.
}
    \label{fig7}
\end{figure}

Figure \ref{fig7} shows boxplots of the total runtime for the 6049-bus network across each division of the Trial 2 dataset. Overall, the concurrent mode provides faster computation, particularly for the 6049-bus network. The maximum runtime for each network and division for the Threads solver and the concurrent mode, as well as the average computation times, are summarized in Tables~\ref{tab:max_times_threads_sync} and~\ref{tab:avg_speedup_threads_sync}. In all cases, \textcolor{black}{the} concurrent mode reduces both the average computation time and the maximum runtime.

The concurrent mode exhibits similar behavior in the Trial 3 dataset. Results are not included here for the sake of brevity.

\subsection{GPU Solver Performance: Default \\vs Tuned Settings in Concurrent Mode}
\label{secGPU_tuned}

In addition to comparing Threads and GPU solvers in concurrent mode, we analyzed the performance of the GPU solver under default and tuned configurations in concurrent mode. We focused on the 6049-bus system, as it represents the most challenging case among the test networks. The concurrent strategy ensures that once one solver converges, the other is immediately terminated, thereby reducing overall computation time and leveraging the strengths of both solvers.

For the tuned GPU solver, we adapted several key parameters to improve convergence, as summarized in Table~\ref{tab:gpu_tuned_params}. Table~\ref{tab:gpu_default_tuned} presents the number of scenarios solved first by the GPU solver for the 6049-bus system in each division under both default and tuned settings for the Trial 2 and Trial 3 datasets. The table shows that tuning the GPU solver significantly improves its performance, particularly in Trial 2, where the tuned solver wins 20 scenarios compared to only 6 under the default configuration. In Trial 3, the performance improvement is more moderate, with the tuned solver solving 10 scenarios first versus 8 for the default setting. This demonstrates that GPU parameter tuning can complement the concurrent strategy, increasing the likelihood that the GPU solver converges first and reduces overall computation time. The recorded speed-up in average computation time is consistent with the results in Table~\ref{tab:avg_speedup_threads_sync}, but it is notably higher in Division 1. For example, the speed-up reaches 5.2 when using two GPU solvers in concurrent mode, compared to 2.86 when using the GPU and Threads solvers together in concurrent mode.

\textcolor{black}{We also verified that the default and tuned GPU configurations produced identical objective values across all evaluated scenarios. Specifically, the objective-value ratio between the tuned and default GPU configurations was exactly 1 for all test cases. This indicates that the parameter tuning improves the convergence performance of the GPU solver without affecting the quality of the resulting solutions.}

\begin{table}[h!]
\caption{\textcolor{black}{Maximum runtimes for the Threads- and Concurrent-based solvers for the Trial 2 dataset.}}
\centering
\small
\setlength{\tabcolsep}{4pt}
\begin{tabular}{|l||l|c|c|}
\hline
\textbf{Network} & \textbf{Division (D)} & \multicolumn{2}{c|}{\textbf{Maximum Runtime [min]}}  \\
\cline{3-4}
 & & Threads & Concurrent  \\
\hline
\hline
4224-bus & D1 & 1.436 & 0.536 \textcolor{blue}{$\downarrow$} \\
         & D2 & 2.023 & 1.296 \textcolor{blue}{$\downarrow$} \\
         & D3 & 2.101 & 1.378 \textcolor{blue}{$\downarrow$} \\
\hline
6049-bus & D1 & 87.26 & 27.35 \textcolor{blue}{$\downarrow$} \\
         & D2 & 119.61 & 29.00 \textcolor{blue}{$\downarrow$} \\
         & D3 & 119.97 & 42.12 \textcolor{blue}{$\downarrow$} \\
\hline
6717-bus & D1 & 5.350 & 3.735 \textcolor{blue}{$\downarrow$} \\
         & D2 & 17.64 & 16.78 \textcolor{blue}{$\downarrow$} \\
         & D3 & 15.67 & 12.51 \textcolor{blue}{$\downarrow$} \\
\hline
\end{tabular}
\label{tab:max_times_threads_sync}
\end{table}

\begin{table}[h!]
\caption{\textcolor{black}{Average runtimes for Threads versus Concurrent solvers and the achieved speed-up for the Trial 2 dataset.}}
\centering
\small
\setlength{\tabcolsep}{4pt}
\begin{tabular}{|l||l|c|c|c|}
\hline
\textbf{Network} & \textbf{Division (D)} & \multicolumn{2}{c|}{\textbf{Avg. Runtime [min]}} & \textbf{Avg.} \\
\cline{3-4}
 & & Threads & Concurrent & \textbf{Speed-up} \\
\hline
\hline
4224-bus & D1 & 1.398 & 0.457 & \textcolor{blue}{$3.06\texttt{x}$} \\
         & D2 & 1.952 & 1.205 & \textcolor{blue}{$1.62\texttt{x}$} \\
         & D3 & 1.846 & 1.144 & \textcolor{blue}{$1.61\texttt{x}$} \\
\hline
6049-bus & D1 & 16.955 & 5.929 & \textcolor{blue}{$2.86\texttt{x}$} \\
         & D2 & 36.366 & 11.364 & \textcolor{blue}{$3.20\texttt{x}$} \\
         & D3 & 35.131 & 18.984 & \textcolor{blue}{$1.85\texttt{x}$} \\
\hline
6717-bus & D1 & 5.285 & 3.698 & \textcolor{blue}{$1.43\texttt{x}$} \\
         & D2 & 8.565 & 8.359 & \textcolor{blue}{$1.02\texttt{x}$} \\
         & D3 & 9.189 & 6.713 & \textcolor{blue}{$1.37\texttt{x}$} \\
\hline
\end{tabular}
\label{tab:avg_speedup_threads_sync}
\end{table}

\subsection{Solver Diagnostics}
Although failures are rare, examining the scenarios in the Trial 2 dataset where either the GPU or Threads solvers fail reveals a consistent pattern. In these cases, the crossover step (which follows the barrier or PDHG phase) terminates with a large dual violation, requiring additional simplex steps to clean the solution. However, the simplex iterations progress very slowly in these scenarios and often stall around $10^3$, which exceeds the convergence tolerance. As a result, the solver is unable to converge within the 120-minute time limit we set. Across all 116 scenarios (57 from Trial 3 and 59 from Trial 2), such failures are extremely rare: both solvers fail in only 2 cases (scenarios 31 and 33 for the 6049-bus network in Division 2), as summarized in Table~\ref{T2_fail}. Interestingly, when the number of threads is reduced to 24 in Section~\ref{secG}, scenario 31 is likely solved in less than 30 minutes due to changes in the solution process, leaving only one scenario (scenario 33 from Division 2 for the 6049-bus system) without a solution within the 120-minute limit. 

\subsection{Impact of Thread Count on \\Barrier Convergence}
The number of threads can influence the convergence path of the barrier algorithm. Multi-threading may alter the order of updates and factorizations, affecting where convergence slows or stalls. The effect is problem-dependent: some instances benefit from more threads, while others achieve more stable convergence with fewer threads.

\subsection{Potential for Additional Acceleration}
\label{sec:additional_parallel}

Beyond the strategies presented above, additional acceleration can be achieved by running multiple instances of the Threads solver (e.g., default and tuned) in parallel, as well as multiple GPUs simultaneously. By testing several Threads solver instances with different settings and executing multiple GPU solvers with varying configurations, overall computation time can be further reduced, terminating all solvers once one converges.

Solver performance can be further enhanced by tuning presolve and factorization parameters. In particular, the Gurobi parameters \texttt{Aggfill}, \texttt{Predual}, and \texttt{Prepasses} control fill during presolve aggregation, dual reductions, and the number of presolve passes, respectively. These adjustments reduce the size and complexity of the linear systems solved at each iteration, improving CPU and GPU performance, and when combined with parallel execution of multiple solver instances, provide a complementary strategy to reduce total computation time. Additional tuning for the Threads solver can be achieved by solving the relaxation part using only the barrier method instead of the default combination of simplex, dual simplex, and barrier methods.

\subsection{\textcolor{black}{Scalability}}

\textcolor{black}{To evaluate the scalability of the GPU-based solver beyond the 6717-bus network, additional experiments were conducted on the largest available networks in Trial 3, Division 1: the 8316-bus and 23643-bus networks. Since only one instance is available at each network size, these results provide an initial assessment of scalability rather than a comprehensive scaling analysis.\
\\
For the 8316-bus network, the GPU-based solver achieved a speedup of 1.22$\times$, reducing the total computational time from 10.8 min to 8.8 min. The improvement primarily comes from reduced computational time for solving the relaxed problem using PDHG and for the subsequent crossover phase.\
\\
For the 23643-bus network, the time spent solving the relaxed problem and performing crossover decreased from 78.2 min with the CPU-based barrier method and crossover to 61.6 min with GPU-based PDHG and crossover, a reduction of approximately 27\%. However, the overall computational time decreased from 177 min to 162 min, resulting in a speedup of 1.09$\times$. The smaller overall speedup is primarily due to the substantial computational time spent in the branch-and-bound phase, which is not accelerated by the GPU implementation.\
\\
These results demonstrate that the proposed GPU-based approach remains applicable to networks substantially larger than 6717 buses. However, the limited number of available instances prevents drawing a definitive scaling relationship with network size.}

\section{Conclusion}

This work evaluates a GPU-accelerated framework for solving large-scale unit commitment problems using the Primal–Dual Hybrid Gradient algorithm. Numerical results on high-fidelity, million-scale UC instances with 4,224-, 6,049-, and 6,717-bus systems demonstrate speedups of up to $3.2\times$ without compromising solution quality. In addition, a concurrent GPU--CPU framework is proposed, where both solvers are executed simultaneously and computation is terminated as soon as either solver converges. This hybrid strategy leverages the complementary strengths of GPU and multi-threaded CPU implementations. The results show that this concurrent mode consistently reduces both average and worst-case computation times across all test systems. These tests highlight the potential of GPU-based optimization to improve the scalability and efficiency of power system operations, with further gains expected as GPU hardware and algorithms continue to advance. 

\begin{table}[t]
\centering
\small
\caption{Tuned GPU solver parameters used for the 6049-bus system.}
\begin{tabular}{|l||l|}
\hline
\textbf{Parameter and Description} & \textbf{Value} \\ \hline
\hline
Absolute feasibility tolerance & $10^{-5}$ \\
Relative tolerance checks & Disabled \\ 
Convergence tolerance & $10^{-4}$ \\ 
Optional iteration limit & $100,000$ \\ \hline
\end{tabular}
\label{tab:gpu_tuned_params}
\end{table}

\begin{table}[t]
\caption{\textcolor{black}{Number of scenarios solved first by each solver on the GPU under default and tuned settings in concurrent mode for each division of the 6049-bus system.}}
\centering
\small
\begin{tabular}{|l||l|l|l|l|}
\hline
\textbf{Division} & \multicolumn{2}{c|}{\textbf{Trial 2}} & \multicolumn{2}{c|}{\textbf{Trial 3}} \\
\cline{2-5}
 & Default & Tuned & Default & Tuned \\
 \hline
\hline
D1 & 1 & 8 & 4 & 2 \\
D2 & 2 & 6 & 3 & 3 \\
D3 & 3 & 6 & 1 & 5 \\
\hline
\hline
\textbf{Total} & 6 & 20 & 8 & 10 \\
\hline
\end{tabular}
\label{tab:gpu_default_tuned}
\end{table}

Artificial intelligence–based tools assisted in refining the readability and organization of this manuscript.

\end{document}